\documentclass[a4paper, 10pt]{article} 

\usepackage[centertags]{amsmath}
\usepackage{amsmath}
\usepackage[dutch,british]{babel}
\usepackage{amsfonts}
\usepackage{amssymb} 
\usepackage{amsthm}
\usepackage{newlfont}
\usepackage{color}
\usepackage{subfig}
 \usepackage{bbm}
 \usepackage{float}
 \usepackage{import}
  \usepackage{epsfig}
  \usepackage{authblk}

\usepackage{stmaryrd}
\usepackage{mathrsfs}
\usepackage{pstricks,pst-node}
\usepackage{fancyhdr}
\usepackage{graphicx}
\usepackage{fancybox}
\usepackage{geometry}
 \usepackage{psfrag}
 
 \usepackage{multirow}

\theoremstyle{plain}
  \newtheorem{theorem}{Theorem}[section]

\theoremstyle{definition}

\theoremstyle{remark}

\numberwithin{equation}{section}

 \newcounter{smallarabics}
\newenvironment{arabicenumerate}
{\begin{list}{{\normalfont\textrm{\arabic{smallarabics})}}}
  {\usecounter{smallarabics}\setlength{\itemindent}{0cm}
  \setlength{\leftmargin}{5ex}\setlength{\labelwidth}{4ex}
  \setlength{\topsep}{0.75\parsep}\setlength{\partopsep}{0ex}
   \setlength{\itemsep}{0ex}}}
{\end{list}}

\newcounter{smallroman}

\newcommand{\ben}{\begin{arabicenumerate}}
\newcommand{\een}{\end{arabicenumerate}}

\DeclareMathOperator{\Tr}{Tr}

\renewcommand{\Re}{\mathrm{Re}\, }

\newcommand\otimesal{\mathop{\hbox{\raise 1.6 ex
  \hbox{$\scriptscriptstyle\mathrm{al}$}
\kern -0.92 em \hbox{$\otimes$}}}}
\newcommand\oplusal{\mathop{\hbox{\raise 1.6 ex
  \hbox{$\scriptscriptstyle\mathrm{al}$}
\kern -0.92 em \hbox{$\oplus$}}}}
\newcommand\Gammal{\hbox{\raise 1.7 ex
\hbox{$\scriptscriptstyle\mathrm{al}$}\kern -0.50 em $\Gamma$}}
\renewcommand\i{\mathrm{i}}

   \let\ep=\epsilon

 \let\la=\lambda

  \let\La=\Lambda

\newcommand{\caL}{{\mathcal L}}

\newcommand{\caO}{{\mathcal O}}

\newcommand{\bbC}{{\mathbb C}}

\newcommand{\bbR}{{\mathbb R}}

\newcommand{\bbZ}{{\mathbb Z}}

\newcommand{\opunit}{\text{1}\kern-0.22em\text{l}}

\newcommand{\norm}{ \|}
\newcommand{\str}{ |}

\newcommand{\e}{{\mathrm e}}
\newcommand{\ed}{{\mathrm e}}

\newcommand{\dd}{{\mathrm d}}

\newcommand{\beq}{ \begin{equation} }
\newcommand{\eeq}{ \end{equation} }
\newcommand{\bet}{ \begin{theorem} }
\newcommand{\eet}{ \end{theorem} }

\begin{document}

\title{Step Heat Profile in localized chains}

\author[1]{Wojciech De Roeck}
\author[2]{Abhishek Dhar}
\author[3]{Fran\c cois Huveneers}
\author[1]{Marius Sch{\"u}tz}
\affil[1]{Instituut voor Theoretische Fysica, KU Leuven, 3001 Leuven, Belgium}
\affil[2]{Universit{\'e} Paris-Dauphine, PSL Research University, CNRS, CEREMADE, 75016 Paris, France}
\affil[3]{International Centre for Theoretical Sciences, TIFR, Hesaraghatta, Bengaluru 560089, India}

\maketitle

\begin{abstract}

We consider two types of strongly disordered one-dimensional Hamiltonian systems coupled to baths (energy or particle reservoirs) at the boundaries: strongly disordered quantum spin chains and disordered classical harmonic oscillators. These systems are believed to exhibit localization, implying in particular that the conductivity decays exponentially in the chain length $L$.  We ask however for the profile of the (very slowly) transported quantity in the steady state. 
We find that this profile is a step-function, jumping in the middle of the chain from the value set by the left bath to the value set by the right bath. The width of the step grows not faster than $\sqrt{L}$. This is confirmed by numerics on a disordered quantum spin chain of 9 spins and on much longer chains of harmonic oscillators.
In the case of harmonic oscillators, we also observe a drastic breakdown of local equilibrium at the step, resulting in a chaotic temperature profile.

\end{abstract}

\vspace{0.5cm}

\section{Introduction}

\begin{figure}[h]
\centering
{\small
        \def\svgwidth{0.7\textwidth}%
        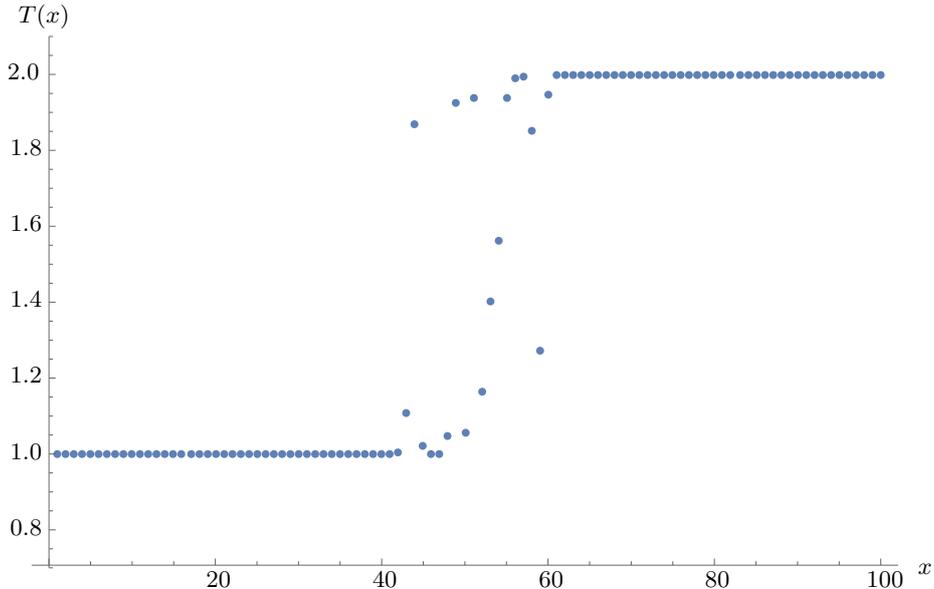
}
\caption{Example of $T(x)$-profile for a disordered harmonic chain of length $L=100$  with distinct boundary regions dominated by the nearest reservoirs with temperatures $T_1=1$ and $T_L=2$ and a fluctuating middle region. See Section \ref{sec: coupled oscillators} for the precise setup; coupling between oscillators is $g=0.1$, oscillator-bath coupling $\lambda \to 0$.}\label{fig: TempProfileExample}
\end{figure}

The study of boundary-driven nonequilibrium systems remains an intriguing problem in statistical mechanics. While 'normal transport', i.e.\ governed by Fourier's or Fick's law, is ubiquitous, the absence of normal transport is also a robust feature, occurring in a range of one-dimensional models \cite{lepri2003thermal,dhar2008heat}.
In this paper, we want to add another behavior to the collection, namely that of localized systems where the temperature (or chemical potential) profile is described by a sharp step-function: if a localized system of length $L$ is placed between two reservoirs at temperatures $T_1$ and $T_2$, then  $T(x)$, with $0 \leq x \leq L$, satisfies
$$T(Ly)\to \begin{cases}   T_1 &   0\leq y < 1/2   \\[2mm]  T_2 &   1/2 < y \leq 1      \end{cases}, \qquad \text{as $L\to \infty$} 
$$
see e.g.\ Figure \ref{fig: TempProfileExample} for some numerical data.
Localized systems have been investigated intensively in the past years, the appeal coming partially from their most striking properties: absence of thermalization and absence of conduction, see e.g.\ \cite{anderson1958absence,imbrie2016many,pal2010many,basko2006metal,gornyi2005interacting}
 for original works and \cite{nandkishore2014many} for a review.    One often distinguishes between Anderson localization as a property of non-interacting particles or modes, and Many Body Localization (MBL) where the Hamiltonian includes interactions between the modes or particles, but apart from this difference in setup, the localization phenomena are analogous.  
We discuss disordered classical pinned harmonic oscillators (Anderson localization) in Section \ref{sec: coupled oscillators} and strongly disordered quantum spin chains (MBL) in Section \ref{sec: mbl}.  In both cases, we provide theoretical and numerical support for the step profiles.   One could also consider a setup without boundary driving, but starting from an infinite nonequilibrium state which becomes thermal when moving away from the origin, with different temperatures to the left and right. In this case one also finds step profiles, but without width of order localization length, i.e.\ not growing with $L$. This setup can be referred to as a quench, see \cite{ros2016remanent,vasseur2015quantum,serbyn2014quantum,abdul2016entanglement} for details, and we do not discuss it in the present paper. 

\paragraph{Intuitive picture via eigenmodes/LIOM's}

A system of harmonic oscillators is best described via its eigenmodes.
A successful characterization of localized quantum spin chains is via the existence of a full set of (quasi)local integrals of motion (LIOMs) \cite{serbyn2013local,huse2014phenomenology, imbrie2016many}, that can be considered as generalizations of eigenmodes: the Hamiltonian contains only action coordinates, but it is not simply a sum of action coordinates.
Concretely, the LIOM theory says that there exists so-called l-spin-operator that are similar to the original spin operators, but they 1)commute with each other and with the full Hamiltonian and 2) are quasilocal instead of local. The full Hamiltonian has interaction terms between the $\ell$-spins, but these interaction terms are not capable of transporting locally conserved quantities.  So the LIOM picture basically amounts to replacing real spins by l-spins. The real spins are of course more explicit objects but the Hamiltonian couples and flips them. The $\ell$-spins are theoretical constructs, but they have the advantage that the Hamiltonian cannot flip them.

This leads immediately to an explanation of the step profiles that we observe.  The only way heat can flow through the $\ell$-spins is by direct contact with the reservoirs. This direct contact exists because the $\ell$-spins are quasilocal and hence have a (very small) overlap with the boundaries of the system.  This overlap decays exponentially in the distance between the $\ell$-spin and the boundary and hence the $\ell$-spin is dominated by the nearest bath.  The same reasoning applies for harmonic systems with `LIOM' replaced by `eigenmode'.  For the harmonic chain, we will derive an explicit expression for the temperature profile in the weak coupling limit, see \eqref{lambda 0 profile}, which confirms
this intuitive argument.

The intuitive explanation also shows why there is room for non-local effects and absence of local equilibrium.  The main point is that the effects that in normal systems lead to local equilibrium, namely local equilibration, are absent here and therefore give way to nonlocal effects that, despite being very weak, can manifest themselves in the absence of local forces.   We should immediately add that we did not exhibit the absence of local equilibrium in a robust way, in particular we do not know whether it occurs for interacting localized systems, where our numerics is limited to short chains (up to 9 sites).  However, for the disordered harmonic oscillators, we clearly see that the kinetic temperature in the middle of the chain varies wildly from site to site as long as we don't average over disorder, see again Figure \ref{fig: TempProfileExample}.

\paragraph{Scope of the theory}

How broad is the described phenomenology?  For example, can it also apply to some anharmonic classical systems?
Most authors would agree that there is no genuine localization in interacting classical systems\cite{basko2011weak,oganesyan2009energy}, but it is clear that a trace of localization remains, and this effect was dubbed 'asymptotic localization' by two of the authors, see \cite{huveneers2013drastic,basko2011weak,de2014asymptotic,de2015asymptotic}.  Roughly speaking this means that all transport is non-perturbative in some parameter. 
For example, for a pinned, disordered, anharmonic chain of oscillators, the conductivity 
clearly vanishes as the strength of the anharmonic interactions (governed by some parameter g)
goes to $0$.
 Asymptotic localization is then the statement that the conductivity decays faster than any polynomial in $g$. 

This sheds some light on the question as to whether the step-profile can persist in such systems. 
Assuming that normal transport eventually takes place in generic (i.e.\ anharmonic) classical systems, we are led to the following picture: As soon as transport due to local conduction mechanisms becomes larger than the direct coupling of LIOM's (that also exist here merely in an asymptotic sense) to the reservoirs, the step profile will give way to normal behavior. 
Assuming that, say, the conductivity $\kappa \sim \ed^{-g}$ and the coupling of the middle LIOM's to the bath goes as $\ed^{-L}$, we deduce that normal behavior will set in as soon as $L$ outgrows a length scale that grows as $1/g$. Hence the step profile will be visible for finite sizes only. 

Let us comment that weird temperature profiles have indeed been observed for rotor chains in \cite{iacobucci2011negative,iubini2014boundary}, and this was one of the motivations for the present paper. 
However, we have no good reason to attribute those temperature profiles to finite-size localization effects. First of all, the considered rotor chains are not in the regime where one expects these effects and secondly, there is a satisfactory explanation for these profiles via linear response \cite{iubini2016coupled}. 

\section*{Acknowledgments}
This work grew out of discussions of F. Huveneers with J.~Lebowitz and D.~Huse. We are most grateful to both of them. Moreover F.~H.\@ thanks J.~Lebowitz for his kind invitation at IAS (Princeton), where this work started.
F. H. thanks the IAS (Princeton), the ANR grant JCJC, and the CNRS InPhyNiTi Grant
(MaBoLo) for financial support.   AD acknowledges  support   from the Indo-Israel joint research project No. 6-8/2014(IC) and from the  French Ministry of Education through the grant ANR (EDNHS).  W.D.R. and M.S. are thankful to the DFG (German Research Fund) and the InterUniversity Attraction Pole DYGEST (Belspo, Phase
VII/18) for funding. 
 

\section{Coupled harmonic oscillators}  \label{sec: coupled oscillators}
We consider a system of harmonic oscillators, situated at a finite set of sites $\Lambda \subset \bbZ^d$ (later we will restrict attention to a chain).  Each site $x \in \La$ carries coordinates $(p_x,q_x) \in \bbR^2$ and we write $(p,q) \in \bbR^{2 \str \La \str}$.  The Hamiltonian is a a quadratic function of $(q,p)$: 
\beq
\mathcal H (q,p) \; = \;  \frac{1}{2}   \langle p | p\rangle + \frac{1}{2}  \langle q |ÊH |Êq\rangle, \qquad H = V - g\Delta, \qquad \langle f|Êh \rangle = \sum_{x\in \Lambda} f_x \overline{h}_x
\eeq
for some $0<g<1$. 
The matrix $V$ (onsite-potential) is diagonal. Its entries $V_x$ are i.i.d. random variables satisfying $V_x \ge v > 0$ for all $x\in \Lambda$.  The matrix $\Delta$ is the discrete Laplacian modeling the harmonic interaction between oscillators.  Since $g\geq 0$ and the spectrum of $-\Delta$ is in $[0,2d]$, $H$ is strictly positive (all eigenvalues are strictly positive).  In the language of coupled oscillators, one says that the  system is 'pinned' and this will mean that it is free
of many intriguing phenomena arising in the unpinned harmonic chain where the conservation of momentum protects the zero mode against localization, see \cite{dhar2001heat,casher1971heat}.  
Boundary conditions are irrelevant for our purposes, but for concreteness, we have taken $\Delta$ with free boundary conditions. 
Physically, the bath acts on (a subset of) the boundary of $\Lambda$.   We single out a nonempty boundary set $\partial \Lambda $ and we couple each of the sites $x $ in $\partial \La$ to an independent bath at temperature $T_x$. This coupling is modeled by Langevin dynamics. 
The equations of motion are: 
\begin{align} 
\dd q_x (t) \; &= \; p_x(t) \dd t,   \qquad x \in \Lambda \label{eq q} \\
\dd p_x (t) \; &= - (Hq)_x (t) \dd t,  \qquad x \in \Lambda \setminus \partial \Lambda \label{eq q no boundary} \\ 
\dd p_x (t) \; &= - (Hq)_x (t) \dd t - \lambda p_x (t)  \dd t + \sqrt{2\lambda T_x}  \dd b_x(t),    \qquad x \in  \partial \Lambda    \label{eq p boundary}
\end{align}
where we have taken for simplicity the coupling strength $\lambda$ the same for all baths.   The $b_x(t)$ are independent Brownian motions. As is well-known, the relation between the friction term $- \lambda p_x (t)  \dd t$ and the noise term $\sqrt{2\lambda T_x}  \dd b_x(t)$ is fixed by requiring that the system is detailed balance at temperature $T_x$ if only the bath at $x$ were present.   As we will take different temperatures, the full dynamics does not satisfy detailed balance and a nontrivial nonequilibrium steady state (NESS) is expected. 
\paragraph{Covariance of NESS}
Let us rewrite the equations of motion (\ref{eq q}-\ref{eq p boundary}) in the more abstract form 
\begin{equation}\label{Gaussian}
\dd X (t) \; = \; A X(t) \dd t + S \dd B(t) 
\end{equation}
with $X(t) = (q(t),p(t)) \in \bbR^{2|\Lambda|}$, with
$$A \;  = \;
\left(
\begin{array}{cc}
0 & 1 \\ -H & -\lambda \sum_{x\in \partial \Lambda} |Êx \rangle \langle x |Ê
\end{array}
\right), \qquad 
S \; = \; 
\left(
\begin{array}{cc}
0 & 0 \\ 0 & \sum_x \sqrt{2\lambda T_x} |Êx \rangle \langle x |Ê
\end{array}
\right) .
$$
For almost all realization of the disorder $A$ is diagonalizable and the real part of all eigenvalue of $A$ is strictly negative, provided $\lambda >0$, see \cite{casher1971heat} for more details.    This also implies that  \eqref{Gaussian} eventually reaches a unique steady state that we denote by  $\langle \cdot \rangle_{\mathrm{ness}}$. 
This invariant measure is Gaussian and is thus characterized by its covariance matrix: 
$$ \langle X X^\dagger \rangle_{\mathrm{ness}} \; = \; \int_0^\infty \dd s\, \ed^{As} SS^\dagger \ed^{A^\dagger s}.$$
For sufficiently small $\lambda$, spectral perturbation theory guarantees that the imaginary part of the eigenvalues is nonzero, and, since the matrix $A$ is real, it follows that eigenvalues come in pairs, related by complex conjugation. Hence we will label them by $k \in  \{ \pm1, \ldots, \pm\str \Lambda \str \}$ and we write the spectral decomposition $A = \sum_k E_k P_k$, with $\Re(E_k)< 0$.  
Hence
\begin{equation}\label{covariance}
\langle X X^\dagger \rangle_{\mathrm{ness}} \; = \; \sum_{k,l} \int_0^\infty \dd s\,  \ed^{E_k s } P_k SS^\dagger P_l^\dagger \ed^{E_l^* s} \; = \; -\sum_{k,l} \frac{1}{E_k + E_l^*} P_k SS^\dagger P_l^\dagger.
\end{equation}
Let us specialize this formula for the case of a one-dimensional chain $\Lambda =\{1,\ldots, L\}$ and $\partial \Lambda=\{1,L\}$.  For convenience we write  $\str x \rangle \in \bbC^{2L} $ for the base vector given by $q_y=0,p_y= \delta_{y,x}, y=1,\ldots,L$.  Then we can write 
$$\langle p_x^2 \rangle_{\mathrm{ness}} \; = \; \big\langle \langle x |ÊX X^\dagger | x \rangle \big\rangle_{\mathrm{ness}}$$ 
with $1 \le x \le L$. 
In that case
$$ SS^\dagger \; = \;  2 \lambda T_1 |Ê1 \rangle \langle 1 |Ê + 2 \lambda T_L |ÊL \rangle \langle L | .$$
Hence
\begin{equation}\label{profile lambda positive}
\langle p_x^2 \rangle_{\mathrm{ness}}   =  - 2\lambda T_1 \sum_{k,l} \frac{1}{E_k + E^*_l} \langle x |ÊP_k |Ê1 \rangle \langle 1 | P_l^\dagger |Ê x \rangle
- 2\lambda T_L \sum_{k,l} \frac{1}{E_k + E^*_l} \langle x |ÊP_k |ÊL \rangle \langle L | P_l^\dagger |x \rangle.
\end{equation}
This expression  in eq.~\eqref{profile lambda positive} can be used as a starting point for numerics, but to get some insight, it is useful to consider the weak coupling limit $\la \to 0$.
\subsection{The weak coupling limit $\lambda \to 0$} \label{sec: derivation weak coupling}
In this limit, the contact with the baths is very weak. One can obtain an expression that involves only the temperatures and the eigenstates of the isolated Hamiltonian and this expression allows for a better interpretation.  This limit was also considered in the same setting in \cite{matsuda1970localization,lepri2003thermal,dhar2012nonequilibrium}.
 
Let us first diagonalize the operator $A$ at $\lambda = 0$.   First, we diagonalize $H$ as 
\begin{equation}\label{eigenstates H}
H|\psi_k\rangle = \omega_k^2 \lvert\psi_k\rangle \qquad \text{with $\omega_k > 0$ and $k=1,\ldots, L$}
\end{equation}
The (conventional) notation $\omega_k^2$ is justified since $H>v$ is positive and bounded below, due to the  choice $v>0$.
Then, the eigenvalues of $A$  at $\lambda = 0$ are
\begin{align*}
& E_k =  \i \omega_k\quad\text{with}\quad E_{-k}=-E_k,\quad k=\pm 1,\ldots, \pm L.
\end{align*}
Note that they are labeled by $k=\pm1,\ldots, \pm L$. 
The spectral projections are 
$$P_k=\frac{1+\omega_k^2}{2\i \omega_k}\lvert \phi_k\rangle \langle \widetilde \phi_k\rvert, \qquad \text{with} \quad   \phi_{ k}=
 \frac{1}{\sqrt{1+\omega_k^2}}\left( 
\begin{array}{c}
\psi_k \\ 
 \pm \i \omega_k \psi_k
\end{array}
\right),  \quad  \widetilde\phi_{ k}=
 \frac{1}{\sqrt{1+\omega_k^2}}\left( 
\begin{array}{c}
\mp\i \omega_k\psi_k \\ 
  \psi_k
\end{array}
\right)  $$

We see that in the $\lambda \to 0$ both numerator (because of the explicit factor $\lambda$) and denominator  (because $E_k=-E^*_{-k}$)  of  the formula \eqref{profile lambda positive} go to zero.  
The spectrum is discrete and non-degenerate, hence for $\lambda > 0$ small enough, the eigenvectors/eigenvalues of $A$ can be determined by perturbation theory
(in general this requires $\lambda$ to be small enough compared to the level spacing, i.e.\@ small enough compared to $1/L$). 
Since $\omega_k$ are bounded away from zero, and there is a coefficient $\lambda$ in the r.h.s.\@ of \eqref{profile lambda positive}, only terms with $k= l$ do not vanish in  the limit $\lambda \to 0$. Using the explicit form of $P_k$ above and the first order energy shifts
$$\frac{\dd E_{k}}{\dd \lambda} = -\Tr\bigl(P_k (\lvert 1\rangle\langle 1\rvert+\lvert L\rangle \langle L\rvert)\bigr) =-\frac{1}{2} (|\psi_k(1)|^2 + |\psi_k(L)|^2),$$ we obtain the $\la\to 0$ limit of \eqref{profile lambda positive}: 
\begin{equation}\label{lambda 0 profile}
\langle p_x^2 \rangle_T  \; = \;  \sum_{k=1}^L  |\psi_k(x)|^2 \Big( T_1 \frac{|\psi_k(1)|^2}{|\psi_k(1)|^2 + |\psi_k(L)|^2} + T_L \frac{|\psi_k(L)|^2}{|\psi_k(1)|^2 + |\psi_k(L)|^2} \Big).
\end{equation}
\paragraph{Validity of the limit $\lambda \to 0$.}
This limit will be convenient to develop a heuristic picture, but one may ask whether it makes sense to first take the limit $\lambda \to 0$ and then analyze the temperature profile in the thermodynamic limit $L\to\infty$.   In general, for a non-localized system this does indeed not make sense, for the following reason. Assuming that the system is diffusive, a quantity of energy injected by one of the baths needs a time of order $\tau_{\mathrm{diff}} \sim L^2$ to diffuse towards the other bath. If the coupling is so small that the time $1/\lambda$ needed to inject some energy, becomes larger than $\tau_{\mathrm{diff}}$, then one should not expect to see any trace of the spatial location of the baths, i.e.\ of the place where the energy got injected, left or right.   Hence, in the limit $\la\to 0$, we expect a flat temperature profile.  This claim can certainly be verified in the (stochastic) exclusion process with different chemical potentials.

Let us further test this prediction by considering a $d=3$ system,  with two reservoirs acting on the left and right edge of the cube $\Lambda$.   If the disorder is small enough, one expects to have a normal heat conductor and a profile interpolating linearly between left and right reservoir \cite{chaudhuri2010heat}.  
In the limit $\lambda \to 0$, an expression analogous to eq.~\eqref{lambda 0 profile} can be derived: 
\beq   \label{eq: threed formula} \langle p_x^2 \rangle_T \; = \; \sum_k |\psi_k(x)|^2 \times \frac{\sum_{y\in \partial \Lambda}T_y |\psi_k(y)|^2}{\sum_{y\in \partial \Lambda}|\psi_k(y)|^2}.
\eeq
with $T_y = T_{cold}$ for $y$ on the left edge, and $T_y = T_{hot}>T_{cold}$ for $y$ on the right edge.  
It is hard to imagine how \eqref{eq: threed formula} can give rise to a linear profile. Indeed, if we assume the modes $\psi_k$ to be completely delocalized, then \eqref{eq: threed formula} suggests that the $T(x)$ should be roughly $x$-independent, in accordance with the above intuition. 

\begin{figure}[H]
\centering
{\small
        \def\svgwidth{0.7\textwidth}%
        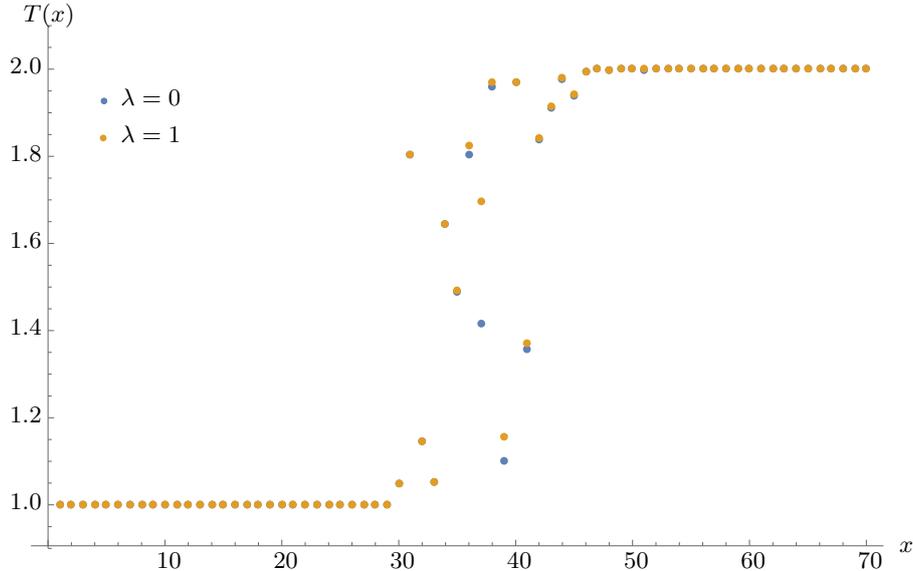
}
\caption{Example of $T(x)$ profile for coupling $g=0.1$, chain length $L=70$, and different values of $\lambda$. There are distinct boundary regions dominated by the nearest reservoirs with temperatures $T_1=1$ and $T_L=2$ and a fluctuating middle region. Typically, the difference between $\lambda=1$ and the limit $\lambda\rightarrow 0$ is noticeable only in this middle region. }\label{fig: TempProfileExamplelambda}
\end{figure}

How is then for a localized system?  The main difference is that here there is no mechanism for the system itself, i.e.\ without assistance of the baths, to transfer energy. In a certain sense  $\tau_{\mathrm{diff}}=\infty$ here and hence the above problem never arises: the only mechanism for transfer of energy is due to the fact that the baths speak directly with the same modes, and this happens via the exponential tails of the localized modes. 
Let us say this on a more technical level by looking directly at the derivation of the weak coupling limit above. There we see that the $\la\to 0$ limit is  similar to the situation $\la>0$ (independent of $L$) whenever spectral perturbation theory for the eigenstates and energies of the matrix $A$ is justified. A numerical test is shown in Figure \ref{fig: TempProfileExamplelambda}.  This is exactly what we expect in a localized system, whereas for delocalized (chaotic) systems spectral perturbation theory is not valid: a small local perturbation tends to mix the eigenstates.

Finally, we note that our derivation of the $\la\to 0$ is not very intuitive. In the appendix, we give a much more transparent derivation that yields also an effective equation. This derivation is very well known in quantum mechanics, where it is the standard way to obtain a master equation for a weakly coupled open system, see \cite{davies1974markovian,spohn1978irreversible}.

%
%

\paragraph{Step profile from eq.~\eqref{lambda 0 profile}.}
Assuming the simplest possible cartoon for the eigenstates, we find that eq.~\eqref{lambda 0 profile} leads to a step profile. 
Let us indeed assume that $\psi_k$ is centered around a site $x_k$ so that 
\beq   \label{eq: form of wavefunctions}
 |\psi_k(x)|^2 \sim   \exp{\left(-2 \frac{|x-x_k|}{\xi(\omega^2_k)}\right)}.
 \eeq
 where $\xi(\omega^2)$ is the (frequency-dependent) localization length.
Inserting this ansatz in eq.~\eqref{lambda 0 profile} and using $\sum_k \str\psi_k(x)\str^2=1$ we find the step profile: 
\beq 
\langle p_{x}^2 \rangle_T = \begin{cases} T_1 &   \text{ if $x < (L-1)/2 $}   \\[2mm]    T_L   &  \text{ if $x > (L-1)/2$}    \end{cases}  \qquad \qquad  \text{up to a term of $\caO(\ed^{-|x-(L-1)/2|/\xi})$}
\eeq
where here $\xi$ is understood as the supremum over $\xi(\omega^2)$. 
This approximation captures well the reason why we see a step profile, but it does not describe the transition region accurately. In particular, if one takes \eqref{eq: form of wavefunctions} above literally, it seems that the width of the step is of order $\xi$, i.e.\ not growing with $L$. 
\paragraph{Step profile smoothed by fluctuations}

We now refine \eqref{eq: form of wavefunctions} to take into account the effect of disorder fluctuations. Since the decay of localized eigenfunctions away from their center is governed by the local disorder and the eigenvalue (cf.\ transfer matrix formalism), it is reasonable to put forward that, for each mode $k$
$$
 |\psi_k(1)|^2 \approx  \ed^{-2 \frac{x_k-1}{\xi_k} +  \eta_{1,k}\sqrt{ k-1 } }, \qquad    |\psi_k(L)|^2 \approx  \ed^{-2 \frac{L-x_k}{\xi_k} +  \eta_{L,k}\sqrt{ L-k} }
$$
where $\xi_k=\xi(\omega^2_k)$ and where $\eta_{1,k},\eta_{L,k}$ are mean-zero random variables, not growing with $L$, and tending to Gaussians as $L\to \infty$.   For fixed $k$ sufficiently far away from  $1$ and $L$,  $\eta_{1,k},\eta_{L,k}$ can be considered independent, as they depend on the disorder on the left, respectively, right of the localization center $k$.   Let us now look back at \eqref{lambda 0 profile} and compare the two weights
$$
\alpha(1,k)=  \frac{|\psi_k(1)|^2}{|\psi_k(1)|^2 + |\psi_k(L)|^2}, \qquad \alpha(L,k)= \frac{|\psi_k(L)|^2}{|\psi_k(1)|^2 + |\psi_k(L)|^2}.
$$
satisfying $\alpha(1,k)+\alpha(L,k)=1$. These  weights determine which of the two reservoirs is felt by the mode $k$. Their ratio is
\beq  \label{eq: log ratio}
 \log  \frac{\alpha(1,k)}{\alpha(L,k)} =  \log  \frac{|\psi_k(1)|^2}{ |\psi_k(L)|^2} =   2 \frac{(L+1-2x_k)}{\xi_k} + \beta_{k}, \qquad \beta_k:= \eta_{k,1}\sqrt{x_k-1}- \eta_{k,L}\sqrt{ L-x_k},
\eeq
Here $\beta_k$ is a random variable with mean zero and variance of order $\sqrt{L}$ (for what follows it is not necessary to try to be more precise). 
Hence if $k$ lies far away from the center: $\str L+1-2x_k \str  \gg C \sqrt{L}$, then of course the weight of the nearest reservoir dominates. This just confirms the intuition above.  
In the middle region, i.e.\ when  $\str L+1-2x_k \str  \leq c \sqrt{L}$, then the zero-mean random variable dominates \eqref{eq: log ratio} and we see that there is an appreciable probability for $ \log  \frac{\alpha(1,k)}{\alpha(L,k)}$ to have either sign.  Moreover, we see that the expectation of $ (\log  \frac{\alpha(1,k)}{\alpha(L,k)})^2$ is of order $L$  itself, so it is likely that one of the weights is close to $1$ whereas the other one is close to $0$.  
This concludes the analysis of the weights for a single mode $k$, let us now consider the different modes together.  

For fixed $L$, 
we expect the variables $\beta_{ k}$ and $\beta_{k'}$ to be strongly correlated when their energies $E_k$ and $E_{k'}$ are close to each other, but this correlation diminishes as   $E_k,E_{k'}$ are separated.  This means that, for two  modes $k,k'$ within the middle region defined above, it is well possible that one of them is dominated by the left bath and the other by the right bath. 
This leads to the following picture for 
$$T(x) = \sum_{k}  \str\psi_k(x)\str^2  (T_1 \alpha(1,k)+ T_L \alpha(L,k)). $$
For concreteness, we assume that we are at strong disorder, $\sup_{\omega} \xi(\omega^2) \ll 1$, so that in the above formula we may approximate $ \str\psi_k(x)\str^2=\delta_{x_k,x}$. Then, we expect that within the middle region $T(x)$ wildly oscillates between the values $T_1$ and $T_2$, and outside of the middle region it assumes the temperature of the nearest bath. 
\beq  \label{eq: theory profile}
T(x)= \begin{cases} T_1  &  x \leq L/2-C\sqrt{L} \\ T_2  &  x \geq L/2+C\sqrt{L}  \\
\text{randomly either $T_1$ or $T_2$} \qquad & L/2-C\sqrt{L}  < x< L/2+C\sqrt{L}
\end{cases}
\eeq
with $C \sim \xi$.
  So in the middle region we indeed see a spectacular breakdown of local equilibrium, as the kinetic temperature jumps between nearest neighbours. 
This is confirmed by numerics, see Figures \ref{fig: TempProfileExample} and \ref{fig: TempProfileExamplelambda}.

\begin{figure}[H]
\centering
{\small
        \def\svgwidth{0.7\textwidth}%
        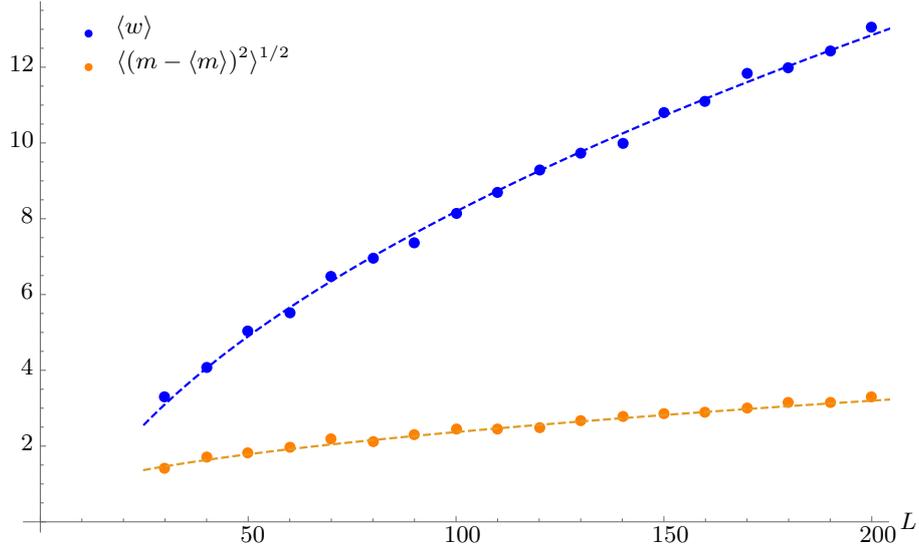
}
\caption{Mean step width $w$ and deviation of the step position $m$ in the $T(x)$ profile averaged over $1000$ realizations of the disorder for different chain lengths $L$, with bath temperatures $T_1=1$ and $T_L=2$, and coupling $g=0.1$. Both quantities grow sublinearly in  $L$. Here $w$ and $m$ are defined as follows: for every profile $T(x)$, let $x_-$ be the first site in the chain where $T(x_{-})> T_1+0.3(T_L-T_1)$ and $x_+$ the last site where $T(x_+)<T_L-0.3(T_L-T_1)$, then the step width can be quantified as $w=x_+-x_-+1$ and the step's midpoint as $m=(x_+ + x_-)/2$. The broken lines are fitted functions of the form $T(x)=a+ b\sqrt{x}$.  
}\label{fig: WandX}
\end{figure}

\begin{figure}[H]
\centering
{\small
        \def\svgwidth{0.7\textwidth}%
        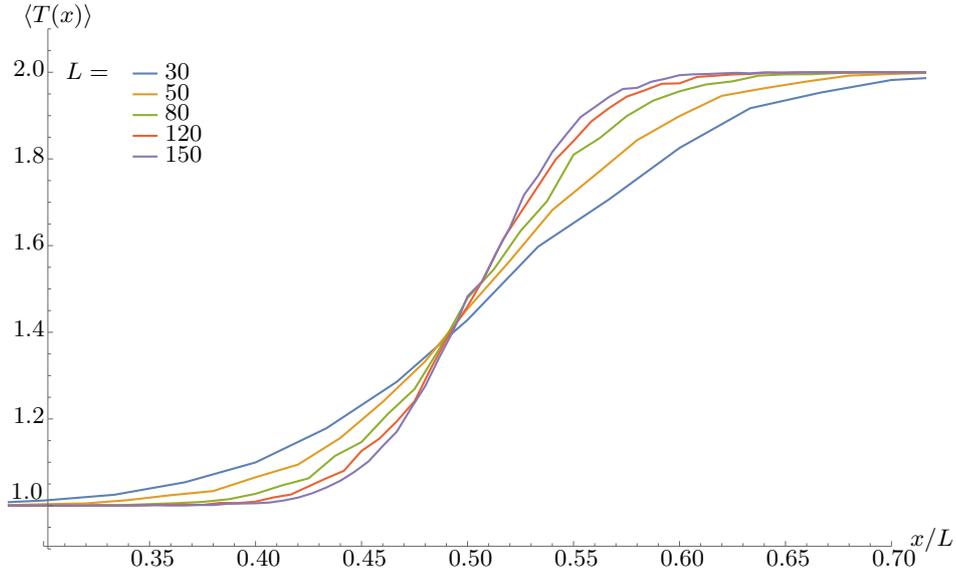
}
\caption{Mean $T(x)$ profile averaged over $1000$ realizations of the disorder for coupling $g=0.1$, with bath temperatures $T_1=1$ and $T_L=2$, and different system sizes $L$. The width of the step in the averaged profile grows slower than linearly in $L$.}\label{fig: Templar}
\end{figure}

\begin{figure}[H]
\centering
{\small
        \def\svgwidth{0.7\textwidth}%
        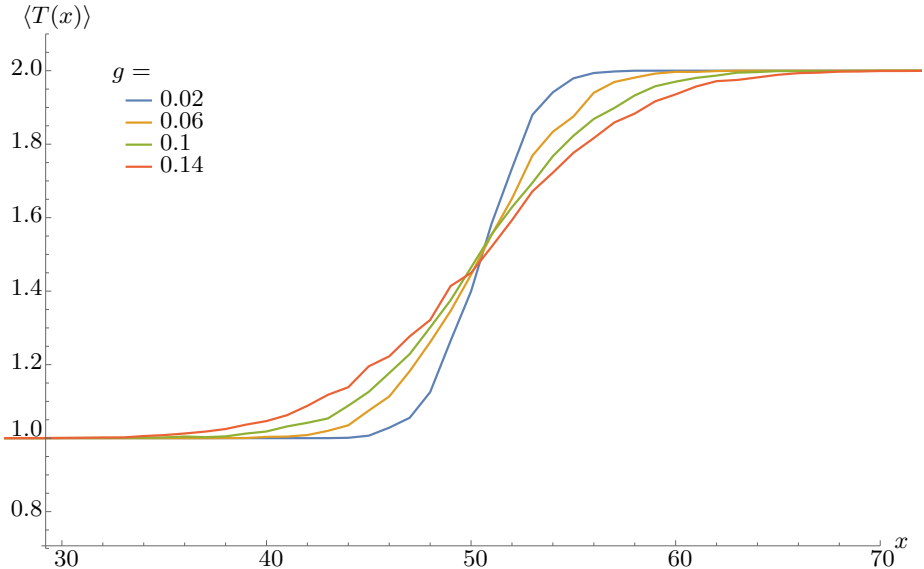
}
\caption{Mean $T(x)$ profile averaged over $1000$ realizations of the disorder for chain lengths $L=100$, with bath temperatures $T_1=1$ and $T_L=2$, and different couplings $g$. The step is becoming sharper for smaller values of $g$, consistent with the behaviour of $C$ in eq.\ \eqref{eq: theory profile}}\label{fig: AvTempG}
\end{figure}

Upon averaging $T(x)$ over disorder, the profile in the middle region is of course smoothed and we get a smoothed step, with theory predicting its width to be  $\sqrt{L}$. We numerically verified that both the average step width and fluctuations of the step position increase indeed slower than linearly in the system size, see Figure \ref{fig: WandX}, though the $\sqrt{L}$ behavior could not be confirmed conclusively.
Averaged profiles for different system sizes and couplings  are shown in Figures \ref{fig: Templar} and \ref{fig: AvTempG}.

\section{Disordered spin chain in the MBL phase}  \label{sec: mbl}

Our model is the disordered XXZ chain
\beq
H_{XXZ} =  \sum_{i= 1}^{L-1}  J( {S}^{+}_i  {S}^{-}_{i+1} +hc)  + V{S}^{z}_i {S}^{z}_{i+1}   + \sum_{i= 1}^L h_i  {S}^z_i
\eeq
where the  operators ${S}_i^{+},{S}_i^{-},{S}_i^{z}$ act on site $i$ as copies of the standard spin operators
$$
{S}^z=  \left( 
\begin{array}{cc}   1& 0  \\  
0  & -1 
\end{array}
\right), \qquad     {S}^+=  \left( 
\begin{array}{cc}   0& 1  \\  
0  & 0 
\end{array}
\right),    \qquad     {S}^-=  \left( 
\begin{array}{cc}   0& 0  \\  
1 & 0 
\end{array}
\right). 
$$
The fields $h_i$ are i.i.d.\ random variables drawn uniformly from the interval  $[-h,h]$.  The parameter
$$
\epsilon \equiv  J/(2h) 
$$
quantifies the (inverse) disorder strength, whereas the term $V{S}^{z}_i {S}^{z}_{i+1}$ breaks integrability of the model. We will always assume 
$$
\ep \ll 1, \qquad     V/h \ll 1. 
$$
Apart from the energy, also the $z-$ magnetization 
$M^z=\sum_x {S}^z_x$ is locally conserved. It will be most convenient for us to study the profile of the magnetization, rather than energy or temperature, in a NESS. 
The above model can also be cast as a model of interacting fermions. Via the Jordan-Wigner transform, $H$ can be related to
\beq
H_{F} = J \sum_{i= 1}^{L-1}  ( {a}^{+}_i  {a}^{-}_{i+1} +hc)  + 4Vn^{z}_i n_{i+1}   + 2\sum_{i= 1}^L h_i  n_i
\eeq
with $a_i,a^\dagger_i$ fermion operators and number operators $n_i= a^\dagger_i a_i$.  In this representation, the additional conserved quantity is the total fermion number $N= \sum n_i$. 
The precise correspondence between the spin and fermion operators is that $H$ is unitarily conjugated to $H_F -4V N  + 2V(n_1+n_L)$. The term $-4VN$ commutes with $H_F$ and the term $2V(n_1+n_L)$ merely shifts $h_{1,L} \to h_{1,L}+2V$. 
For the analysis that follows, we remain mostly in the spin picture.

The significance of $\ep$ for perturbation theory is seen as follows: The  term ${S}^{+}_i  {S}^{-}_{i+1} +hc$  flips spins $\str \uparrow \downarrow \rangle \Leftrightarrow \str \downarrow \uparrow \rangle$. The energy difference $\Delta E$ between eigenstates of the $H(J=0)$ Hamiltonian that are connected via such a spin-flip is  $\Delta E= 2\str h_{x+1}-h_x \str +\mathcal{O}(V)$,  and the modulus of the matrix elements is $J$. Perturbation theory in $J$ can in principle apply if $J \leq \Delta E$.  This latter condition is satisfied  typically if $\ep \ll 1$, given that $V \ll h$.  Of course, this reasoning is only conclusive for two spins, but it is now widely accepted that in a certain sense (see below) perturbation theory applies for arbitrarily large chains\cite{imbrie2016many,pal2010many}. Numerics \cite{luitz2015many} set the critical value at $1/\epsilon \approx 7.2$: at smaller $\epsilon$ perturbation theory works for all eigenstates, and there is 'full localization', to be precisely defined in Section \ref{sec: diagonalization}. Many authors predict that at somewhat larger values of $\epsilon$ localization still persists, but then not at all energy densities, i.e.\ a ``many-body mobility edge''. We will however not be concerned with this regime and we stick to 'full localization'. 
%

\subsection{Diagonalization of the Hamiltonian} \label{sec: diagonalization}
In our opinion, the most intuitive and useful characterization \cite{serbyn2013local,imbrie2016many} of MBL is that the Hamiltonian can be diagonalized by a quasilocal unitary  $U$, i.e.\
$$
U H_{XXZ} U^\dagger  =   D  
$$
where $D$ is diagonal in ${S}^z$ basis. 
$$
D=D(\{{S}^z_i\})=  \sum_i  g_i  {S}^z_i   +     g_{i,i+1} {S}^z_i   {S}^z_{i+1} +  g_{i,i+1,i+2} {S}^z_i   {S}^z_{i+1} {S}^z_{i+2}  +\ldots 
$$
where the interaction strengths $g_{i,\ldots,j}$ are functions of the random fields $(h_i)$ and they decay exponentially   $g_{i,\ldots,j} \sim \epsilon^{\str i-j \str}$ in most places, except at resonant spots, see below for more details.
The operator $U$ is quasilocal in a sense that we will explain now. We choose an operator $O_i$ acting at site $i$ and with $\norm O_i \norm=1$.  We describe its transform $U^\dagger O_i U$ by expanding it in local operators.  On general grounds, one can write an expansion of the following type
$$
U^\dagger O_i U =  \sum_{A: I(A) \ni i}    K(A) {S}_{A}
$$
where the sum is over pairs $A=(I(A),\alpha(A))$ with $I(A)$ a  discrete interval $I(A) = \{i_1, i_1+1, \ldots, i_2\}$ with $i \in I(A)$ and $\alpha(A)$ an array of labels $\alpha(A)= (\alpha_{i_1},\ldots,\alpha_{i_2})$
where each $\alpha$ takes values in the set $\{+,-,0,z\}$.  Then ${S}_A=\prod_{i \in I(A)}    {S}_i^{\alpha_i} $ with $S_i^0=1$ and the other $S^{\alpha_i}_i$ as defined above. For consistency, we also require $\alpha_{i_1}\neq 0$ unless $i_1=i$ and the same for $\alpha_{i_2}$.  
Of course, we could as well have chosen another way to parametrize $U^\dagger O_i U$ and as long as we did not fix properties of the coefficients $K(A)$, the discussion is completely general. 
However, we now specify that the coefficients $K(A)$ are quasilocal functions of the fields, i.e.\ they depend primarily on  $h_i, i\in I(A)$ and that they decay as functions of the length $\str I(A)\str$, such that the typical\footnote{Typicality refers here to the random fields $h_i$} behavior of the $K(A)$ is 
\beq  \label{eq: def bigk}
\str K(A) \str \sim  \e^{-\str I(A)\str/\xi}, \qquad \text{with $\xi$ a localization length}
\eeq
 This $\xi$ is mainly determined by $\epsilon$, but $V$ enters as well, in particular when $V$ grows large. When $V=J$ and $\ep \ll 1$, then we find from perturbation theory in $J$ that, roughly speaking, 
$$
1/\xi \simeq   \log (1/\epsilon)
$$
Of course, \eqref{eq: def bigk} is a caricature. First of all, the conservation of $M^z$ has an impact because we can choose the operator $U$ to commute with $M^z$. Let us assume this henceforth and let us consider $O_I= S_i^+$. Then $K(A)=0$ unless $S(A)$ also raises $M^z$ by $2$, i.e.\ unless
$
\str \{ i:  \alpha_i=+   \}  \str -   \str \{ i:  \alpha_i=-   \}  \str =  1 
$.  We will henceforth always implicitly restrict ourselves to $A$ for which conservation laws permit $K(A) \neq 0$.  Then, there can be further systematic deviations from  \eqref{eq: def bigk} when the interaction $V/J$ is very small.  At $V=0$, the system is mapped to non-interacting fermions, which translates to the property (again with $O_I= S_i^+$) that  $K(A)=0$ unless $I(A)=i'$ and  $\alpha_{i'}=+$. 
Furthermore, there are of course deviations from \eqref{eq: def bigk} due to resonant spots, determined by the local disorder, where the spatial decay is absent (or reduced). At such resonant spots, perturbation theory breaks down and the unitary $U$ loses its quasi-local structure. The simplest way to model this is by letting the total decay be a sum of local, fluctuating contributions and splitting into mean and variance, resulting in the following refinement of \eqref{eq: def bigk}:
\beq  \label{eq: fluctuations}
\frac{1}{\str I(A) \str}\log  \str K(A) \str \sim -(1/\xi)  +    \frac{1}{\sqrt{\str I(A) \str}  }  \eta
\eeq
with $\eta$ a random variable whose size does not grow with $A$.  The latter formula is much less explicit than the discussion around eq.\ \eqref{eq: log ratio} since we do not know whether $\eta$ is solely a function of the disorder, or whether it also depends on the labels $\alpha(A)$.

\paragraph{Local Integrals of Motion (LIOM)}

Having diagonalized the Hamiltonian, we now proceed to describe the most striking consequence, namely the existence of a full set of commuting, local, conserved quantities: LIOM's='local integrals of motion'.  

We call these LIOM's $\widetilde{S}^z_i$ where the notational similarity to the $ {S}^z_{i} $ is intentional. One also says that the $ {S}^z_{i} $ describe physical spins or bits whereas the 
$\widetilde{S}^z_i$ describe localized spins or bits, often called $\mathrm l$-spins.  The $\widetilde{S}^z_i$ are defined by 
$$
\widetilde{S}^z_i  \equiv   U^\dagger {S}^z_i U
$$
with $U$ as in the previous section, guaranteeing that $\widetilde{S}^z_i $ are indeed quasilocal. 
Since, by construction, $S^z_i$ commutes with $D(S^z_i)$, we obtain that indeed $[\widetilde S^z_i,H_{XXZ}]=0$ and moreover
\beq  \label{eq: liom ham}
H_{XXZ}=  D(\{\widetilde{S}^z_i\})=  \sum_i  g_i  \widetilde{S}^z_i   +     g_{i,i+1} \widetilde{S}^z_i   \widetilde{S}^z_{i+1} +  g_{i,i+1,i+2} \widetilde{S}^z_i  \widetilde{S}^z_{i+1} \widetilde{S}^z_{i+2}  +\ldots 
\eeq
This is the LIOM-representation of the Hamiltonian. 
We can now also introduce in the obvious fashion other operators acting directly on the LIOM's 
$$
\widetilde{S}^{\alpha}_i  \equiv   U^\dagger {S}^\alpha_i U, \qquad \text{with $\alpha=z,+,-$ as above}
$$
It is then straightforward to see that the $\widetilde S^\alpha_i$ satisfy the same algebra as the $S^\alpha_i$, justifying the terminology of 'spins'. 
Moreover, since at most sites $\widetilde S^\alpha_i-S^\alpha_i =\mathcal{O}(\epsilon)$ by properties of $U$, it is clear that the conservation of the $\widetilde S_i^z$ leads also to nonergodic behaviour for the $S_i^z$ as well. We do not discuss this further since it is extensively explained in many places. 


\subsection{Chain coupled to baths}
Let us consider a system bath-Hamiltonian of the type
\beq
H =  H_{XXZ} +  H_{B_l} +H_{B_r} + \left( {S}^+_1\otimes V_{B_l} + {S}^+_L\otimes V_{B_r} + hc \right)
\eeq
where $H_{XXZ}$ is the Hamiltonian defined above, and $B_l,B_r$ stand for two baths, one on the left and one on the right, with respective Hamiltonians $H_{B_l},H_{B_r}$. The operators of the type  ${S}^+_1\otimes V_{B_l}$ couple the chain to the baths, and we have written for simplicity $H_{XXZ}= H_{XXZ} \otimes 1_{B_l} \otimes 1_{B_r}$, $H_{B_l}  =  1_{\text{chain}} \otimes H_{B_l} \otimes 1_{B_r}$, etc.\ 
We  rewrite this Hamiltonian using the LIOM operators and the unitary $U$: 
$$
H =  D(\widetilde{S}^z) +    H_{B_l} +H_{B_r}+ \left( (U\widetilde{S}^+_1U^\dagger) \otimes V_{B_l} + (U\widetilde{S}^+_LU^\dagger)\otimes V_{B_r} +hc\right)
$$
To get insight into this expression, we need to write $U\widetilde{S}^{\pm}_1U^\dagger, U\widetilde{S}^{\pm}_LU^\dagger$ in terms of the $\ell$-spin operators. 
Strictly speaking, we prescribed in the previous section only how to perform the opposite task, i.e.\ we described the (quasi-)locality of the map $O\mapsto U^\dagger O U$, but obviously the same statements should hold true for the map $O\mapsto U O U^\dagger$, and hence we expect an expansion of the form
$U\widetilde{S}^{\pm}_1U^\dagger = \sum_A K(A) \widetilde S_A$ with $K(A)$ having the properties described above.

Let us focus on the $H_{S,B_l}$, i.e. the coupling of the system to the left bath.  Written in LIOM's this becomes
$$
H_{S,B_l}= ( c_1\widetilde{S}^+_1 +  c_2 \widetilde{S}^z_1\widetilde{S}^+_2 + c_3\widetilde{S}^+_2 +  c_4 \widetilde{S}^+_1\widetilde{S}^z_2+     \ldots)  \otimes V_{B_l}   +hc. 
$$
where we have written here the terms that act up to the second $\ell$-spin and we have used a lighter notation instead of the $K(A)$-coefficients.  Note that only terms enter that raise the magnetization by $+2$, as we already remarked in Section \ref{sec: diagonalization}. 

In this representation, the coupling to the bath is the only term that can flip $\ell$-spins, and thus thermalize the chain. 
Let us look at the $j$ leftmost $\ell$-spins and erase all terms from the Hamiltonian that couple theses $\ell$-spins to the right bath $B_r$. Doing so, we are omitting terms of size $g_r\e^{-(L-j)/\xi}$ where $g_r$ is the strength of the original coupling term to the right bath, i.e.\ of $V_{B_r}$. 
Let us analyze the resulting Hamiltonian: the $j$ leftmost spins are coupled to the left bath by terms that are not smaller than $\e^{-j/\xi}$. Hence these spins should thermalize at temperature $T_{1}$ (of the left bath) in a time of order $(g_l\e^{-j/\xi})^{-1}$, with $g_l$ the strength of $V_{B_l}$. 
If now 
$$
g_l\e^{-j/\xi}  \gg    g_r\e^{-(L-j)/\xi},  \qquad \text{i.e.\ }  \quad   (L/2-j) \gg \xi
$$
then it follows\footnote{To be precise, there is a slight glitch in the above argument: namely the Hamiltonian of the left $l$-spins, which is diagonal in the $\tilde S^z_i$, does depend weakly on the other 
$\ell$-spins via the long-range terms in \eqref{eq: liom ham}.  This means that the equilibrium state for the $j$ leftmost $\ell$-spins depends weakly on the other spins, and those are communicating with the right bath. However, the maximal effect of the $j+1$'th $\ell$-spin is a change of order $\e^{-1/\xi}$ in the Hamiltonian, and  if we restrict attention to the first $j-1$ spins, then this maximal effect is not bigger than $\e^{-2/\xi}$, etc.\  
Hence the conclusion above remains valid.} that these left spins are thermalized by the left bath and assume its temperature, $T_1$. Of course, the same reasoning can be developed for the right half of the chain and so we arrive again at step-like temperature profile.
Fluctuations have not yet been taken into account, and it is clear that the random variable in \eqref{eq: fluctuations} can give rise to a smoothing of the step over a width of order $\sqrt{L}$, after averaging over disorder. 
What happens for typical realizations is not clear to
us: The sizes that we can reach numerically are way too small to make any convincing observation, and the theoretical argument developed for the harmonic chain in Section \ref{sec: coupled oscillators} does not apply since in a many-body system, we cannot longer identify eigenstates with sites. Hence,
our theoretical prediction is here simply
\beq  \label{eq: theory profile mbl}
T(x)= \begin{cases} T_1  &  x \leq L/2-C\sqrt{L} \\ T_2  &  x \geq L/2+C\sqrt{L} 
\end{cases}
\eeq
with $C \sim \xi$.

 \paragraph{Modeling by a Lindblad dynamics}
 
 For our numerical tests, we will rather use the magnetization than the energy as the transported quantity. The natural variable conjugate to magnetization is of course the magnetic field, and equilibrium states are given by $\e^{-\beta (H+hM^z)}$. We will take $\beta=0$ and keep only the potential $\mu \equiv -\beta h$ as a thermodynamic parameter.  To ease intuition, we choose now the fermionic representation so that the conserved quantity is the particle number, and $\mu/\beta$ is the chemical potential. 
 We model the coupling to a bath by a Lindblad semigroup $\rho \mapsto \e^{t \mathcal{L}\rho}$ acting on density matrices $\rho$, with a difference in $\mu$ between left and right bath.  
 This means that we consider the Lindblad operator
 $$
 \caL \rho=  -\i [H_{F}, \rho]  + \caL_{l}\rho +\caL_r\rho
 $$
 with $H_{F}$ the Hamiltonian of the isolated fermion chain,
$$H_F = J\sum_x^{L-1} (a_x^\dagger {a}_{x+1} +hc) + V n_x n_{x+1} +2\sum_{x=1}^L h_x n_x, $$ 
  and 
\beq
\caL_{l}\rho =   \gamma^+_1\big(  {a}^{\dagger}_1 \rho  {a}_1 - (1/2) \{   {a}_1 {a}^{\dagger}_1 , \rho   \}\big)   +     \gamma^-_1\big(  {a}_1 \rho  {a}^{\dagger}_1 - (1/2) \{   {a}^{\dagger}_1 {a}_1 , \rho   \}\big)  
\eeq
 and similarly for the right bath Lindbladian  $\caL_{r}$ but now with rates $\gamma^{\pm}_L$ and with $a_L,a^\dagger_L$ replacing   $a_1,a^\dagger_1$. 
 The associated potentials are 
 $$
 \mu_1=  \log \frac{\gamma^+_1}{\gamma^-_1}, \qquad    \mu_L=  \log \frac{\gamma^+_L}{\gamma^-_L}
 $$
 and we can easily check that at $\mu\equiv \mu_1=\mu_L$, the stationary state is  the product
 $$
 \rho\propto \exp{\left(\mu N \right)}
 $$
Lindblad operators like the above have been used often in many-body physics, also relating to MBL, see \cite{benenti2009charge, levi2016robustness, fischer2016dynamics}, though often one would add as well dephasing in the bulk, which destroys MBL.

 Using the QuTip python package \cite{johansson2013qutip}, we numerically obtained the stationary state (density matrix) $\rho_\mathrm{stat}$ of the above dynamic at $\mu_1 \neq \mu_L$.

\begin{figure}[H]
\centering
{\small
        \def\svgwidth{0.7\textwidth}%
        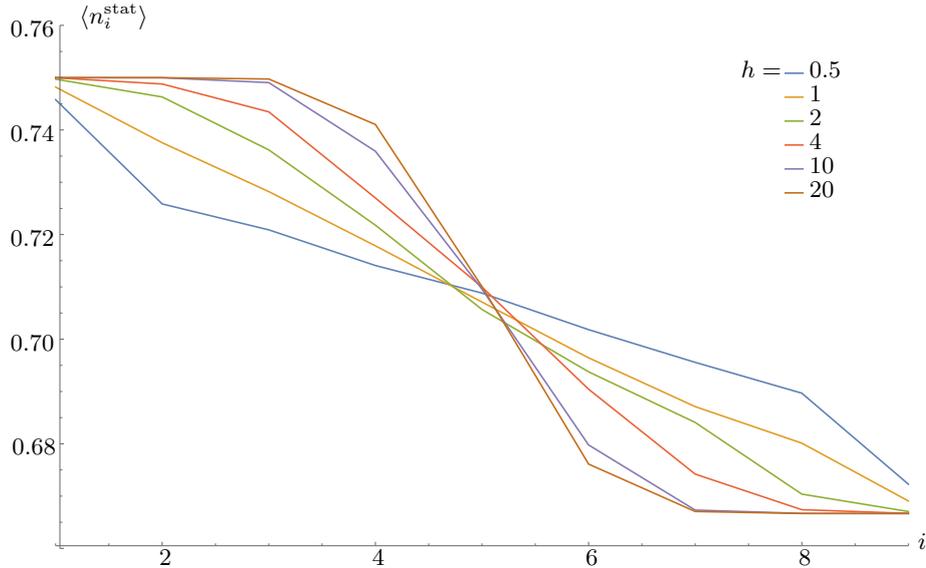
}
\caption{Density profiles $\langle n_x\rangle$ in the stationary state of the Lindblad dynamics for different field strengths $h$ averaged over 50 random realizations of the fields $h_x\in[-h,h]$. The other system parameters are $V=1$, $J=1$, and ($\gamma_1^+$, $\gamma_1^-$, $\gamma_N^+$, $\gamma_N^-$)$=$($3$, $1$, $2$, $1$)}\label{fig: Lindblad}
\end{figure}

\begin{figure}[H]
\centering
{\footnotesize
        \def\svgwidth{\textwidth}%
        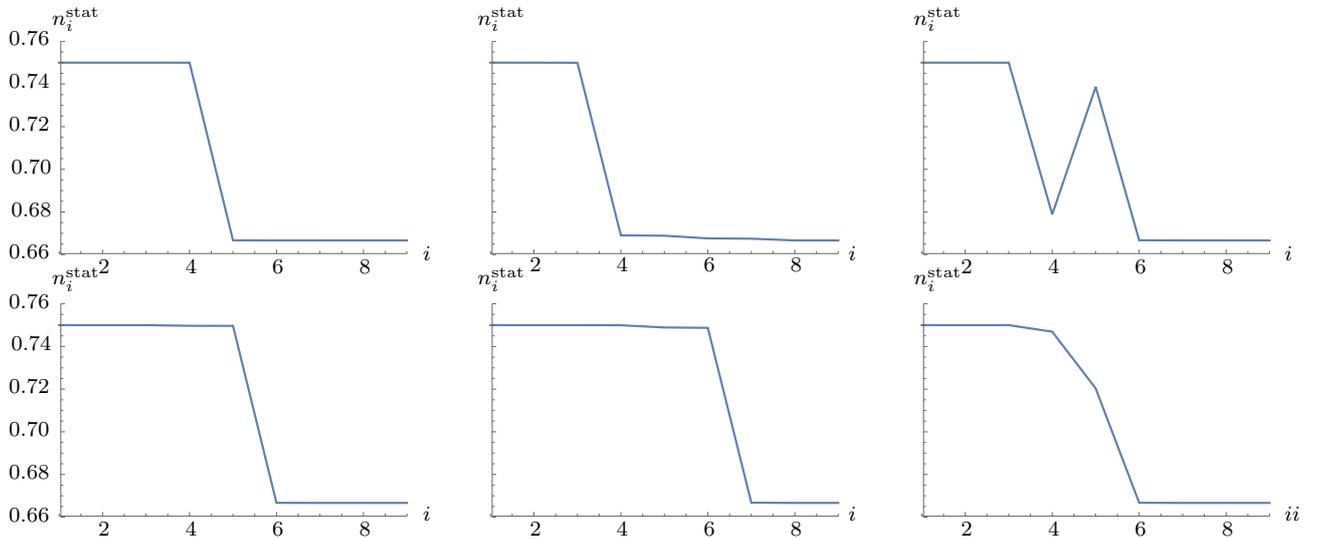
}
\caption{Example profiles $n_i^\mathrm{stat}$ for different realizations with $h=20$. Typically, the step is much sharper than for the averaged profile in Figure \ref{fig: Lindblad}.\label{fig: LindbladExample}}
\end{figure}

 The disorder-averaged stationary profile $ n^{\mathrm{stat}}_i \equiv \langle n_i^\mathrm{stat}\rangle$ for a chain of length $L=9$ is shown in Figure \ref{fig: Lindblad}, for different values of the disorder strength $h$. 
 We see clearly that for higher values of $h$, the profile tends to a step, whereas for a moderate disorder strength $h=1$, the profile is linear. For even smaller disorder strengths, we see a straight profile with the jumps occurring at the boundaries. The latter is likely a finite-volume artefact that derives from the fact that the disorder-free XXZ chain is integrable and has hence ballistic transport.
In any case, the picture concerning step profiles is even clearer if we consider the profile before disorder-averaging, thus discarding the smoothing effect of averaging over the position of the step, see e.g. a few realizations in Figure  \ref{fig: LindbladExample}
 We quantify the sharpness of the non-averaged steps in Table 1. For example, on average $0.072/0.083=87 \%$ of  the total density drop occurs between two neighboring sites for $h=10$. Note that the density profiles need not be monotone in the middle region. This is also visible in the third example in Figure \ref{fig: LindbladExample} and it accounts for the fact that, in the third line of Table 1, the sums of absolute values of nearest neighbor differences exceed the difference between the ends.
\begin{table}[H]
\begin{center}
\begin{tabular}{c||c|c|c|c|c|}
& $h=1$ & $h=2$ & $h=4$  & $h=10$ & $h=20$ \\[2pt]
\hline
$\max_i \lvert  n_i - n_{i+1}\rvert$ \vphantom{$\Bigl($} & 0.026 & 0.051 & 0.061  & 0.072 & 0.071\\
\hline
$\lvert  n_L - n_{1}\rvert$ \vphantom{$\Bigl($} & 0.079 &\multicolumn{4}{c|}{0.083}\\
\hline
$\bigl(\sum_i \lvert  n_i - n_{i+1}\rvert-\max_i \lvert  n_i - n_{i+1}\rvert\bigr)\bigr/(L-2)$ \vphantom{$\Bigl($}& 0.0082 & 0.0075 & 0.0059 & 0.0033 & 0.0026\\
\hline
\end{tabular}
\caption{Quantifying the step shape of the profiles (quantities are averaged over 50 random field configurations). For simplicity we write simply $n_i$ for $n_i^\mathrm{stat}$.}
\end{center}
\end{table}

\begin{figure}[H]
\centering
{\small
        \def\svgwidth{0.7\textwidth}%
        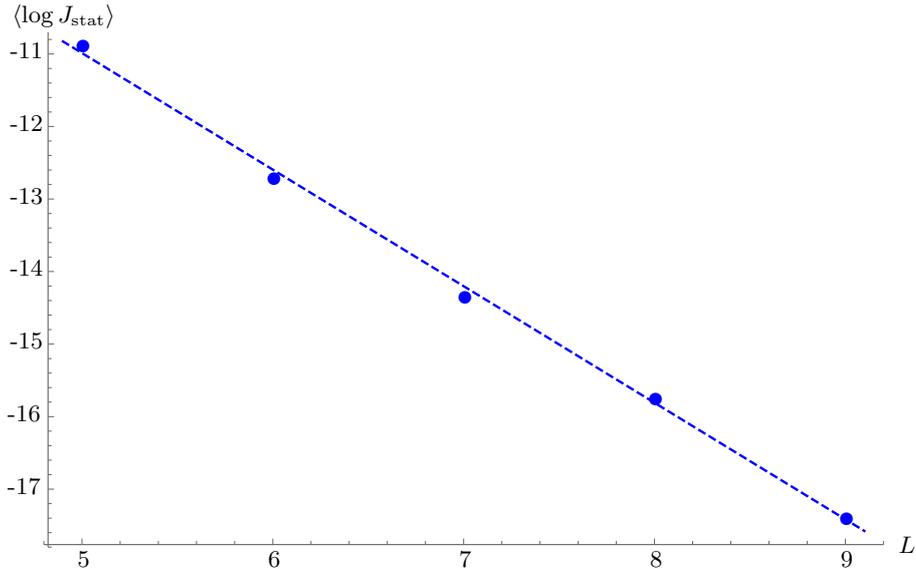
}
\caption{Logarithm of the stationary current $J_\mathrm{stat}$ averaged over $1000$ realizations of the disorder  for chain lengths up to $L=8$ and $100$ realizations for $L=9$. The dashed line is a linear fit.}\label{fig: Current}
\end{figure}

\paragraph{Behavior of the stationary current}
For clarity and completeness, we also comment on the stationary current through the chain. To define it, note that by stationarity and $[H_F,N]=0$, 
$$
0=\partial_t \Tr\bigl(\rho_\mathrm{stat} N\bigr)= \Tr\bigl(\caL_l\rho_\mathrm{stat}  N\bigr)+\Tr\bigl( \caL_r\rho_\mathrm{stat} N\bigr),
$$
The two terms on the right are naturally interpreted as the stationary currents $J_{\mathrm{stat}}$ from the left/right reservoir. 
$$
J_{\mathrm{stat}} \equiv \Tr\bigl(\caL_l\rho_{\mathrm{stat}} N\bigr)=-\Tr\bigl(\caL_r\rho_{\mathrm{stat}} N\bigr)=\gamma_1^+ - (\gamma_1^+ + \gamma_1^-)  n_1^\mathrm{stat},
$$
which depends only on the stationary spin occupation $n_1^\mathrm{stat}$ (or $n_L^\mathrm{stat}$)  at the boundaries.
As already mentioned, the conductivity, and hence the stationary current, decays exponentially with chain length, due to localization. We tested this  numerically up to the accessible length $L=9$, see Figure \ref{fig: Current}. 
In the same spirit, we understand that the mixing time of the process described by the Lindbladian is also exponentially growing with $L$, which is the practical reason why we are restricted to small system sizes in the numerics.

  \bibliographystyle{unsrt}

\bibliography{loclibrary}

\appendix
 
 \section{Alternative derivation of the $\lambda \to 0$ limit}
We propose a second way to derive eq.~\eqref{lambda 0 profile}, i.e.\@ the expression of the NESS profile in the limit $\lambda \to 0$. 
This is somewhat less straightforward but possibly more intuitive. 
We obtain also a little bit more: we derive an effective dynamics in the limit $\lambda \to 0$. 

\paragraph{Equations of motion.}
We massage the equations of motion to arrive at \eqref{new eq motion} below.
First, it is convenient to adopt a matrix notation: 
\begin{align*} 
\dd  |Êq \rangle  \; &= \;  |Êp \rangle  \dd t, \\
\dd |Êp \rangle \; &= - H |Êq \rangle \dd t - \lambda \sum_{x\in \partial \Lambda} p_x |Êx \rangle \dd t + \sqrt{2\lambda} \sum_{x\in \partial \Lambda} \sqrt T_x  \dd b_x(t) |Êx \rangle .
\end{align*}
Next, we introduce an effective time $\tau = \lambda t$, so that the equations become
\begin{align*} 
\dd  |Êq \rangle  \; &= \; \lambda^{-1} |Êp \rangle  \dd \tau, \\
\dd |Êp \rangle \; &= - \lambda^{-1} H |Êq \rangle \dd \tau - \sum_{x\in \partial \Lambda} p_x |Êx \rangle \dd \tau + \sqrt{2} \sum_{x\in \partial \Lambda} \sqrt T_x  \dd b_x(\tau) |Êx \rangle .
\end{align*}
Next we move to the normal modes of the isolated system. Recall \eqref{eigenstates H}. Let us introduce the notations
$$q(k) = \sum_x  \psi_k(x) q_x, \qquad  p(k) =  \sum_x  \psi_k(x) p_x$$
Then the equations of motion can be recast as
\begin{align*} 
\ddÊq(k)  \; &= \; \lambda^{-1}Êp(k)  \dd \tau, \\
\dd p(k) \; &= - \lambda^{-1} \omega_k^2Êq(k)  \dd \tau - \sum_{x\in \partial \Lambda} p_x \psi_k(x) \dd \tau + \sqrt{2} \sum_{x\in \partial \Lambda} \sqrt T_x  \psi_k (x)  \dd b_x(\tau)  .
\end{align*}
It is convenient to pass from the $(q,p)$ coordinates to the $(a^*,a)$ coordinates:  
\begin{align*}
& a(k) \; = \; q(k) + \frac{\i }{\omega_k} p(k), \qquad a^*(k) \; = \; q(k) - \frac{\i }{\omega_k} p(k), \\
& q(k) \; = \; \frac{1}{2} \big( a(k) + a^*(k) \big), \qquad p(k) \; = \; \frac{\omega_k}{2 \i} \big(a(k) - a^*(k)\big). 
\end{align*}
The equations become 
\begin{equation}\label{new eq motion}
\dd a(k) = \frac{- \i \omega_k}{\lambda} a(k) \dd \tau + \frac{\i }{\omega_k} \bigg( - \sum_{x\in \partial \Lambda} p_x \psi_k(x) \dd \tau + \sqrt{2} \sum_{x\in \partial \Lambda} \sqrt T_x  \psi_k (x)  \dd b_x(\tau)  \bigg) .
\end{equation}

\paragraph{Limit $\lambda \to 0$: resonant averaging.}
We now consider the limit $\lambda \to 0$. 
We follow the method used by Dymov in \cite{dymov2015nonequilibrium}.
The first term in the rhs of \eqref{new eq motion} is dominant on short time scales, and if the second term was absent, the solution would be given by $a(k,\tau) = \ed^{-\i \omega_k \tau /\lambda} a(k,0)$. 
This motivates the change of variables
$$ a(k,\tau) = \ed^{-\i \omega_k \tau /\lambda} A(k,\tau).$$
The evolution equation for $A(k)$ is
$$\dd A(k) \; = \; \frac{\i\, \ed^{\i \omega_k \tau /\lambda}}{\omega_k}  \bigg( - \sum_{x\in \partial \Lambda} p_x  \psi_k(x) \dd \tau + \sqrt{2} \sum_{x\in \partial \Lambda} \sqrt T_x  \psi_k (x)  \dd b_x(\tau)  \bigg) .$$
The rhs is oscillating fast at any frequency $\omega_k$ (the spectrum of $H$ is bounded away from 0). 
We exploit this to obtain an expression in the limit $\lambda \to 0$. 
First we analyze the noise term and then the dissipative term. 

\emph{1. Noise term.}
A computation of second moments shows that, in general, 
$$
\sqrt 2 B_{k,x}(t) \; : =   \;  B^{(1)}_{k,x} (t) + \i B^{(2)}_{k,x} (t)  \; : =   \;   \lim_{\lambda \to 0} \sqrt 2 \int_0^t \ed^{-\i \omega_k  s/\lambda} \dd b_{x}(s)$$
where $B^{(i)}_{k,x}, i=1,2$ are standard independent Brownian motions and the limit is in law, jointly for the collection of processes indexed by $k,x,i$. 
As noticed in \cite{dymov2015nonequilibrium}, for a single $k,x$, the noise `bifurcates' in the limit, since two independent copies ($i=1,2$) are generated from a single process.  
There is more here: each Brownian motion `$2|\Lambda|-$furcates', since for each mode $k$ the corresponding noises are independent. 
Hence the noise term in our effective equation will be given by 
\begin{equation}\label{noise term}
\frac{\i \sqrt{2}}{\omega_k} \sum_{x\in \partial \Lambda} \sqrt T_x  \psi_k (x)  \dd B_{k,x}(\tau).
\end{equation}

\emph{2. Dissipative term.}
We work out 
\begin{equation*}
p_x 
\; = \; \sum_k \psi_k^* (x) p(k)
\; = \; \sum_k \psi_k^* (x) \frac{\omega_k}{2 \i} (a_k - a_k^*)
\; = \; \frac{1}{2 \i} \sum_k \psi_k^*(x) \omega_k \big( \ed^{-\i \omega_k \tau/\lambda} A_k - \ed^{\i \omega_k \tau/\lambda} A_k^* \big). 
\end{equation*}
This term will be multiplied by $\ed^{\i \omega_k \tau /\lambda}$. 
Hence, in the limit $\lambda \to 0$, only the term with the factor $\ed^{-\i \omega_k \tau/ \lambda}$ remains. We obtain the dissipation term
\begin{equation}\label{dissipative term}
- \frac{1}{2} \sum_{x\in \partial \Lambda} |\psi_k(x)|^2 A_k .
\end{equation}
In conclusion, we find that the effective equation in the limit $\lambda\to 0$ is given by
$$
dA(k) = \frac{\i \sqrt{2}}{\omega_k} \sum_{x\in \partial \Lambda} \sqrt T_x  \psi_k (x)  \dd B_{k,x}(\tau) - \frac{1}{2} \sum_{x\in \partial \Lambda} |\psi_k(x)|^2 A_k
$$
Since all modes have now been decoupled, it is straightforward to recover the temperature profile \eqref{lambda 0 profile}.

\paragraph{Comments}
This derivation requires the limit  $\lambda\to 0$ to be taken before the thermodynamic limit $\Lambda \to \bbZ^d$.  Indeed, when arguing that the noises corresponding to the different modes $k$ become independent, we relied on $\lambda$ being much smaller than the difference between any two frequencies $\omega_k$. This derivation would hence fail if the $\omega_k$ were to form a continuum. For localized systems, the situation is more subtle: since each site in the chain sees effectively only a few modes, the treatment above becomes correct again. A more careful argument for this was given in Section \ref{sec: derivation weak coupling}.

\end{document}